\documentclass[final,obeyFinal]{IEEEoj_minimalheading}
\pdfoutput=1

\usepackage[utf8]{inputenc}
\usepackage[ngerman,english]{babel}
\usepackage{microtype}

\usepackage[defaultsans,scale=0.95]{opensans}
\usepackage[T1]{fontenc}

\usepackage[pdftex]{graphicx}
\graphicspath{{fig/}}

\usepackage{pgfplots}
\pgfplotsset{compat=1.16}

\usepackage{changes}
\usepackage[acronym]{glossaries}
\usepackage{textcomp}
\usepackage{todonotes}
\usepackage{ifdraft}
\usepackage{sansmath}
\usepackage{booktabs}

\usepackage[binary-units,per-mode=symbol]{siunitx}
\DeclareSIUnit\op{Op}

\usepackage{xcolor}
\definecolor{color0}{HTML}{1F77B4}
\definecolor{color1}{HTML}{FF7F0E}
\definecolor{color2}{HTML}{2CA02C}
\definecolor{color3}{HTML}{D62728}
\definecolor{color4}{HTML}{9467BD}
\definecolor{color5}{HTML}{8C564B}
\definecolor{color6}{HTML}{E377C2}
\definecolor{color7}{HTML}{7F7F7F}
\definecolor{color8}{HTML}{BCBD22}
\definecolor{color9}{HTML}{17BECF}

\usepackage{tikz}
\usetikzlibrary{positioning,arrows.meta,calc,patterns,shapes,backgrounds,fit}

\PassOptionsToPackage{hyphens}{url}\usepackage[hidelinks]{hyperref}
\usepackage[capitalise]{cleveref}

\newcommand{\todoreview}[1]{\textnormal{}}

\title{Demonstrating Analog Inference on the \acrlong{bss2} Mobile System}
\author{%
	Yannik~Stradmann,
	Sebastian~Billaudelle,
	Oliver~Breitwieser,
	Falk~Leonard~Ebert,
	Arne~Emmel,
	Dan~Husmann,
	Joscha~Ilmberger,
	Eric~Müller,
	Philipp~Spilger,
	Johannes~Weis,
	Johannes~Schemmel\IEEEmembership{,~Member,~IEEE}%
}%
\authornote{%
	All authors contributed equally, a detailed description of individual contributions is given at the end of the manuscript.
	The development of the presented system has been funded by the \acrfull{bmbf} under grant number 16ES1127 as part of the \foreignlanguage{ngerman}{\emph{Pilotinnovationswettbewerb \glqq{}Energieeffizientes KI-System\grqq{}}}.
	It is built upon work funded by the EU (H2020/2014-2020: 720270, 785907, 945539 (HBP)), the \foreignlanguage{ngerman}{Lautenschläger-Forschungspreis} 2018 for Karlheinz Meier and the \foreignlanguage{ngerman}{Deutsche Forschungsgemeinschaft} (DFG, German Research Foundation) under Germany’s Excellence Strategy EXC 2181/1-390900948 (Heidelberg STRUCTURES Excellence Cluster).
}
\corresp{%
	All authors are with the Kirchhoff-Institute for Physics, Heidelberg, Germany (e-mail: yannik.stradmann@kip.uni-heidelberg.de).
}

\newacronym{afib}{\mbox{A-fib}}{atrial fibrillation}
\newacronym{asic}{ASIC}{application-specific integrated circuit}
\newacronym{asicab}{\acrshort{asic} adapter \acrshort{pcb}}{\acrlong{asic} adapter \acrlong{pcb}}
\newacronym{bmbf}{BMBF}{German Federal Ministry of Education and Research}
\newacronym{bss2}{\mbox{BSS-2}}{Brain\mbox{ScaleS-2}}
\newacronym{bss1}{\mbox{BSS-1}}{Brain\mbox{ScaleS-1}}
\newacronym{bss}{BSS}{BrainScaleS}
\newacronym{cdnn}{CDNN}{convolutional deep neural network}
\newacronym{cpu}{CPU}{central processing unit}
\newacronym{dfki}{DFKI}{German Research Centre for Artificial Intelligence}
\newacronym{dma}{DMA}{direct memory access}
\newacronym{dram}{DRAM}{dynamic random-access memory}
\newacronym{ecg}{ECG}{electrocardiogram}
\newacronym{fpga}{FPGA}{field-programmable gate array}
\newacronym{i2c}{I\textsuperscript{2}C}{Inter-Integrated Circuit}
\newacronym{ic}{IC}{integrated circuit}
\newacronym{isa}{ISA}{instruction set architecture}
\newacronym{jit}{JIT}{just-in-time}
\newacronym{lvds}{LVDS}{low-voltage differential signaling}
\newacronym{pcb}{PCB}{printed circuit board}
\newacronym{ppu}{\acrshort{simd} \acrshort{cpu}}{\acrlong{simd} \acrlong{cpu}}
\newacronym{relu}{ReLU}{rectified linear unit}
\newacronym{rtl}{RTL}{Register Transfer Level}
\newacronym{simd}{SIMD}{single instruction, multiple data}
\newacronym{snn}{SNN}{spiking neural network}
\newacronym{sodimm}{\mbox{SO-DIMM}}{small outline dual in-line memory module}
\newacronym{sram}{SRAM}{static random-access memory}
\newacronym{vmm}{VMM}{vector-matrix multiplication}
\newacronym{stdp}{STDP}{spike-timing dependent plasticity}
\newacronym{adex}{AdEx}{Adaptive Exponential Integrate-and-Fire}

\usepackage[numbers]{natbib}
\Crefname{equation}{Equation}{Equations}
\Crefname{figure}{Fig.}{Figs.}
\Crefname{tabular}{Table}{Tables}
\Crefname{section}{Section}{Sections}
\Crefname{subsection}{Section}{Sections}
\crefname{equation}{}{}
\crefname{figure}{Fig.}{Figs.}
\crefname{tabular}{Table}{Tables}
\crefname{section}{Section}{Sections}
\crefname{subsection}{Section}{Sections}

\def\BibTeX{{\rm B\kern-.05em{\sc i\kern-.025em b}\kern-.08em T\kern-.1667em\lower.7ex\hbox{E}\kern-.125emX}}
\AtBeginDocument{\definecolor{ojcolor}{cmyk}{0.93,0.59,0.15,0.02}}

\newcommand{\minisection}[1]{\vspace{-1.5\baselineskip}\subsubsection*{#1}}

\begin{document}
		\begin{abstract}
			We present the \acrlong{bss2} mobile system as a compact analog inference engine based on the \acrlong{bss2} \acrshort{asic} and demonstrate its capabilities at classifying a medical \acrlong{ecg} dataset.
The analog network core of the \acrshort{asic} is utilized to perform the multiply-accumulate operations of a \acrlong{cdnn}.
At a system power consumption of \SI{5.6}{\watt}, we measure a total energy consumption of \SI{192}{\micro\joule} for the \acrshort{asic} and achieve a classification time of \SI{276}{\micro\second} per electrocardiographic patient sample.
Patients with \acrlong{afib} are correctly identified with a detection rate of \SI{93.7(7)}{\percent} at \SI{14.0(10)}{\percent} false positives.
The system is directly applicable to edge inference applications due to its small size, power envelope, and flexible I/O capabilities.
It has enabled the \acrlong{bss2} \acrshort{asic} to be operated reliably outside a specialized lab setting.
In future applications, the system allows for a combination of conventional machine learning layers with online learning in spiking neural networks on a single neuromorphic platform.

		\end{abstract}


	\maketitle
	\glsresetall
	\section{Introduction}\label{sec:introduction}

\PARstart{A}{rtificial} neural networks have become an important tool for a broad variety of tasks -- from datacenter to edge applications.
Striving for energy-efficient and fast computation of these networks, a multitude of novel computing architectures have been developed.
Specialized processors either accelerate the processing of artificial \glspl{cdnn} or -- in the field of event-based neuromorphic computing -- follow a neuroscience-oriented approach and implement \glspl{snn}.

Accelerators for \gls{vmm}-based \gls{cdnn} models mostly rely on computational units in the digital domain~\citep{google-edge-tpu,moloney2014myriad2,hickmann2020nervana,reuther2019survey}, although recent analog approaches show very promising performance~\citep{fick2017analog,fick2022analog}.
In agreement with their biological example, event-based neuromorphic systems traditionally utilize analog computational paradigms~\citep{mead88silicon,moradi2018dynaps,benjamin2014neurogrid}, the general availability of modern CMOS process nodes has however boosted the popularity of digital solutions in this field as well~\citep{Khan2008,merolla2014million,frenkel2018online,frenkel2019morphic,frenkel2020convolutional,frenkel2022reckon,mayr2019spinnaker,davies2018loihi}.
Most recently, research of \gls{vmm}, as well as \gls{snn} accelerators has been augmented by the introduction of post-CMOS technologies based on novel materials~\citep{joshi2020accurate,chua2012hodgkin,shastri2021photonics}.

In contrast to aforementioned single-purpose approaches, the \acrlong{bss} neuromorphic architecture combines analog \gls{vmm} with the event-based emulation of \glspl{snn}.
\Gls{bss2} therefore provides a highly configurable computational substrate for research in the combined fields of computer- and neuroscience~\citep{pehle2022brainscales2_nopreprint_nourl,klein2021addressing} and has been shown to achieve beyond-state-of-the-art energy efficiency and classification latency~\citep{cramer2022surrogate}.
Combining potential energy efficiency benefits and online learning capabilities of \glspl{snn} with the high computational power of \glspl{cdnn} on a single \gls{asic} opens up unique opportunities for adaptive inference applications on the edge.
The only other neuromorphic architectures simultaneously supporting rate- and spike-based models are the digital Tianjic~\citep{pei2019tianjic} and MONETA~\citep{kim2022moneta} systems, both however do not enable freely programmable on-chip learning rules.

\begin{figure}
	\center{%
		\includegraphics[width=\linewidth]{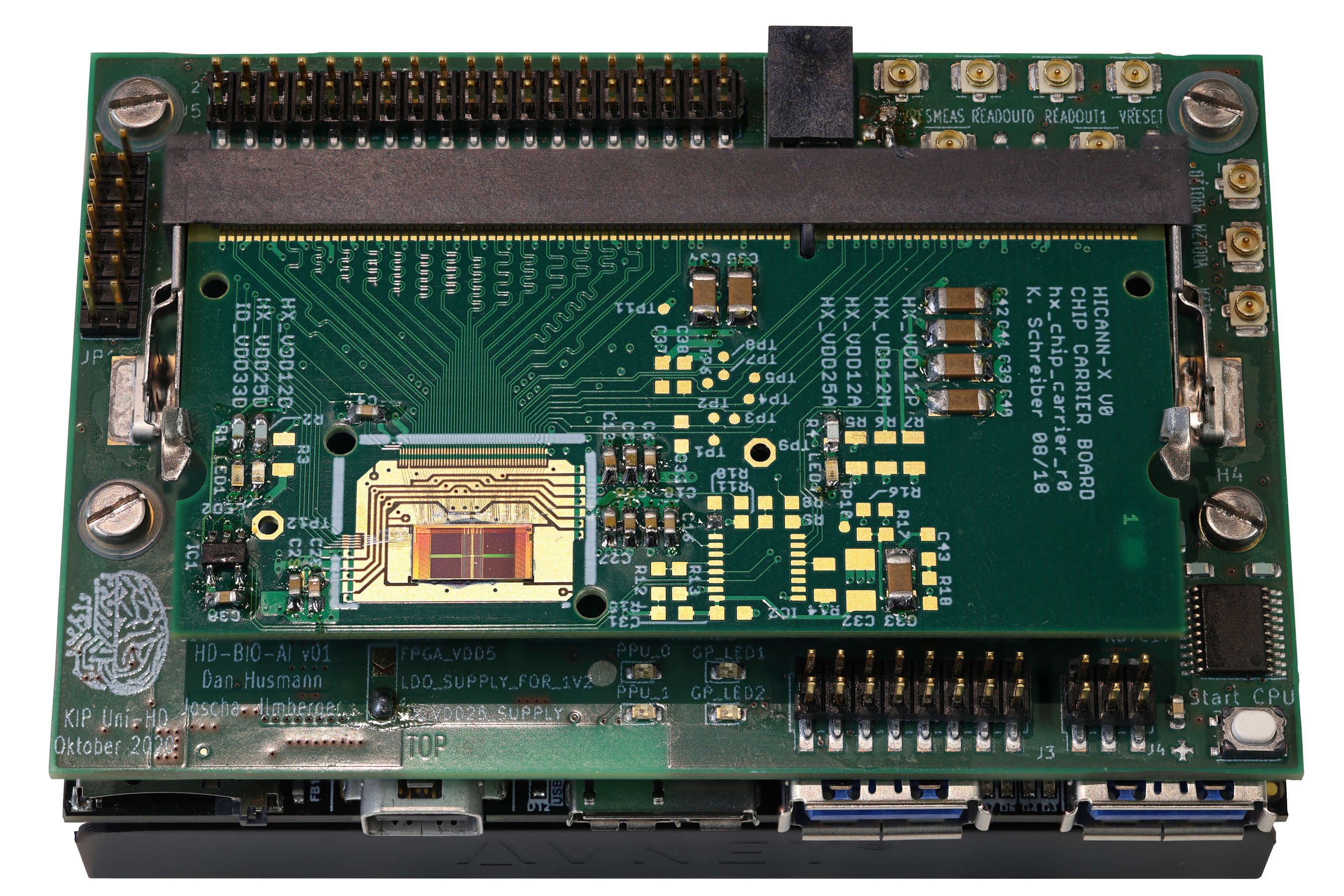}
		\caption{Photo of the \acrlong{bss2} mobile system (from bottom to top): \gls{fpga}-based system controller, \gls{asicab}, \gls{asic} carrier board with the latest \gls{bss2} \gls{asic} directly wire-bonded to the \gls{pcb}.
		The system has the mechanical footprint of a credit card (\SI{84}{\milli\meter}\,\texttimes\,\SI{55}{\milli\meter}) at a height of approximately \SI{40}{\milli\meter}.
		It weighs roughly \SI{155}{\gram} with and \SI{70}{\gram} without the \gls{fpga}'s heatsink respectively.
		\label{fig:system photo}}}
\end{figure}

We now present a highly integrated mobile demonstrator system for the \gls{bss2} architecture (\cref{fig:system overview,fig:system photo}) and showcase the system's capabilities and energy efficiency at the example of \gls{ecg} anomaly classification.
While both, the computation of \glspl{cdnn} and the emulation of \glspl{snn} on \gls{bss2} have already been shown in controlled lab environments~\citep{cramer2022surrogate,weis2020inference}, we can now provide a system that is physically small, has a low power envelope and flexible I/O capabilities.
These previous experiments designed for \gls{bss2} are compatible with the presented mobile platform, the herein presented \gls{ecg} classifier extends the set of applications by a task tailored to edge scenarios.

The design constraints for this system as well as the chosen classification task were motivated by the participation in the independently judged \foreignlanguage{ngerman}{\emph{Pilotinnovationswettbewerb \glqq{}Energieeffizientes KI-System\grqq{}}} by the \acrfull{bmbf}, where it has proven to operate reliably outside controlled lab environments.
This competition posed a challenge to classify \gls{afib} in batches of medical \gls{ecg} recordings with stand-alone edge computing accelerators.
\label{dataset}
The provided dataset consists of \num{16000} traces from the same patient group and has been recorded with two channels only, mimicking the signal quality to be expected from consumer-grade medical wearables.\footnote{Since the dataset contains sensitive patient information it is not publicly available.}
The classification of anomalies in \gls{ecg} time series data is an active field of research where both, classical time series analysis and machine learning-based algorithms compete~\citep{clifford2017af}.

	\section{The BrainScaleS-2 Mobile System}\label{sec:bss2-mobile}

The \gls{bss2} mobile system features a combination of a commercially available \gls{fpga} module and the most recent \acrlong{bss2} \gls{asic}.
The \gls{fpga} contains an embedded CPU which is used for standalone experiment control and I/O\@.
The logic fabric in the \gls{fpga} acts as a memory interface and data format converter for the \gls{asic}.
\begin{figure}
	\center{%
		\includegraphics[width=\linewidth, page=4,viewport=0 0 25cm 14.5cm, clip]{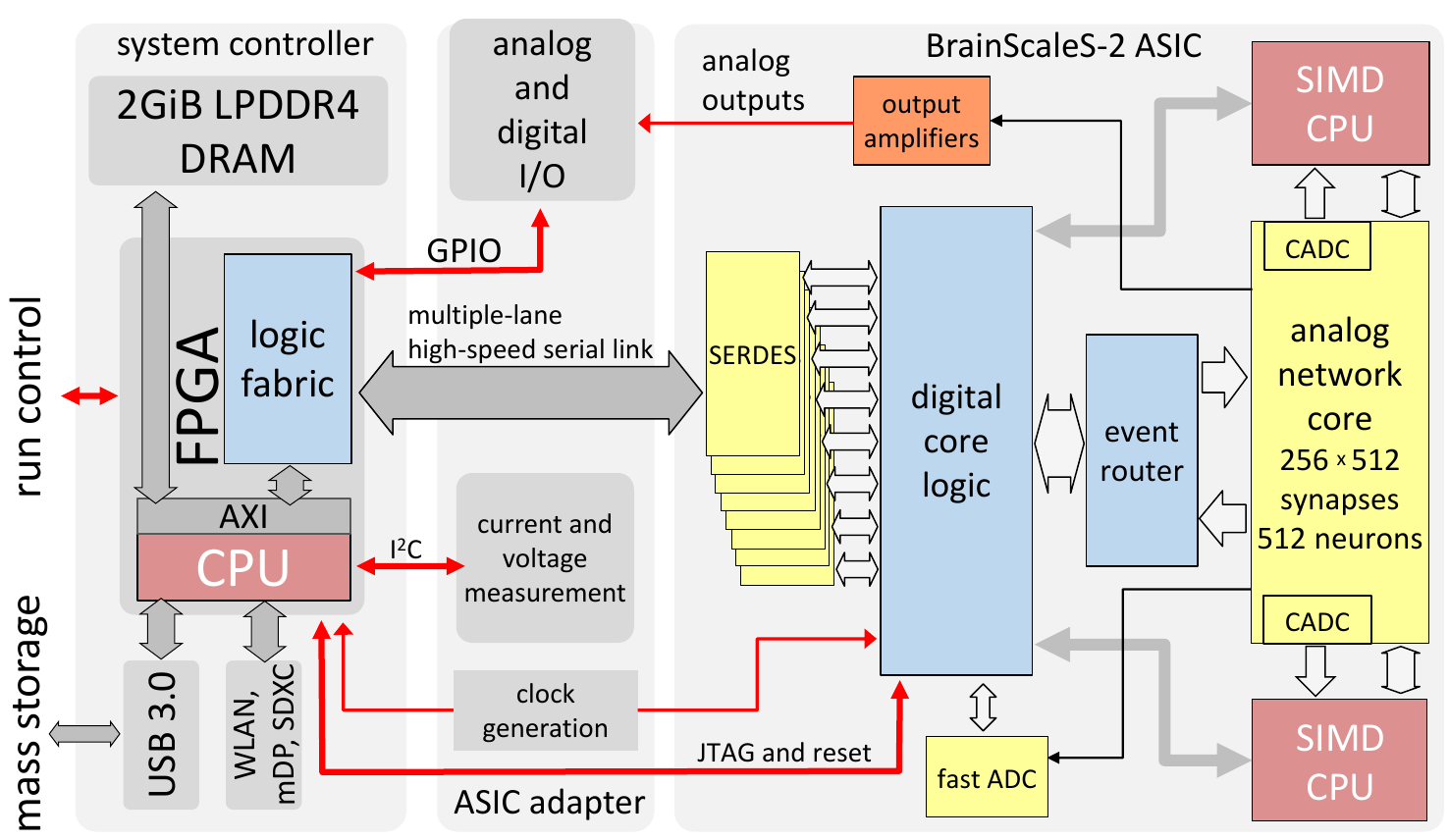}
		\caption{Overview of the \acrlong{bss2} mobile system (from left to right): \gls{fpga}-based controller, \acrshort{asicab} and the \gls{bss2} \gls{asic}.
		\label{fig:system overview}}}
\end{figure}
\Cref{fig:system overview} depicts the three main components of the system:
\begin{itemize}
	\item the \acrlong{bss2} \gls{asic} directly bonded to a carrier board (right),
	\item a custom \acrshort{asicab}, interfacing the \gls{fpga} board to this \gls{asic} carrier board (center),
	\item the system controller, consisting of a low-power \gls{fpga} with an embedded quad-core microprocessor~\citep{xilinx2019zynqultra} and \SI{2}{\gibi\byte} of LPDDR4 \acrshort{dram}, USB~3.0 (device \& host), SDXC, 802.11b/g/n Wi-Fi as well as Bluetooth 4.2 (BLE) communication circuits (left).
\end{itemize}

The described system is the result of a tightly coupled interdisciplinary work ranging from chip design to software engineering and machine learning.
The following sections describe different aspects of the \gls{bss2} mobile system from the perspective of the different technological areas.

\subsection{Neuromorphic ASIC}\label{subsec:neuromorphic-asic}
\begin{figure*}
	\centering
	\includegraphics[width=0.9\linewidth]{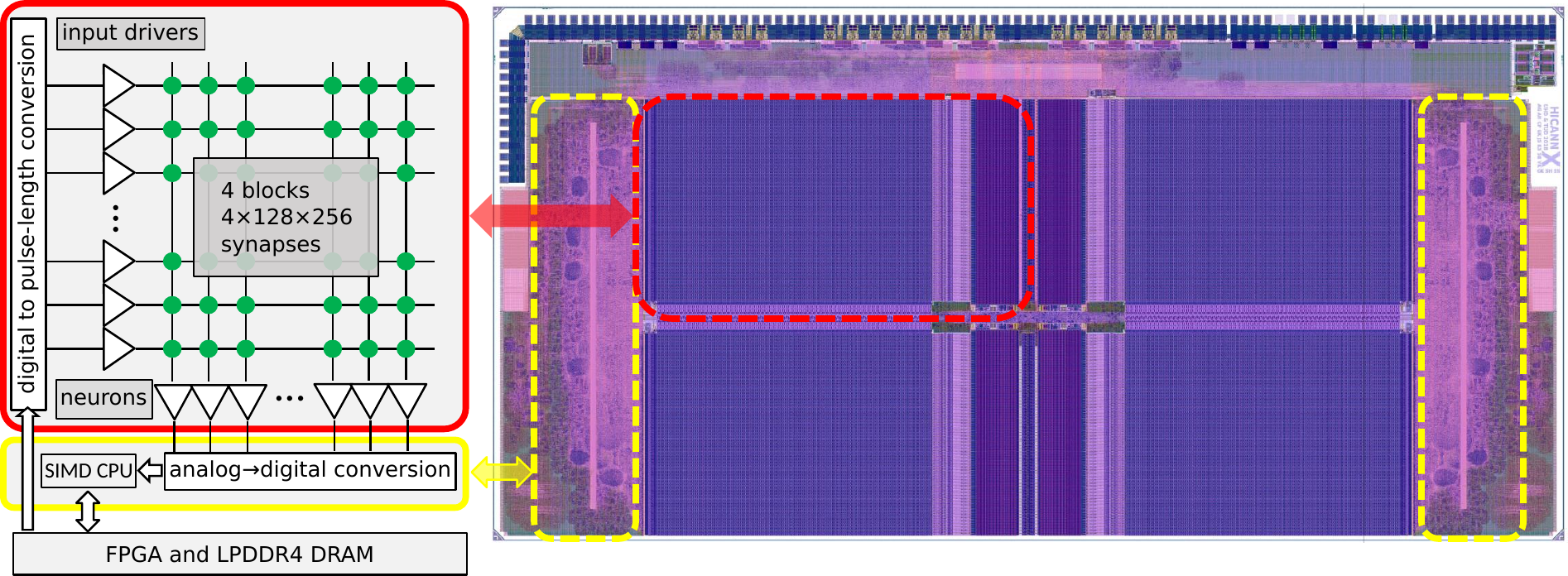}
	\caption{%
		\label{fig:hicann_layout}
		\emph{Left:} internal structure of the \gls{bss2} \gls{asic}.
		The analog network core consists of four quadrants, each containing \num{128} neurons and \num{128}\texttimes\num{256} synapses (red).
		A total of \num{1024} parallel ADC channels allow for readout of various analog parameters by two embedded \gls{simd} processors (yellow).
		\emph{Right:} position of the described functional units on a layout drawing of the \gls{bss2} \gls{asic}.
	}
\end{figure*}
The \gls{bss2} neuromorphic \gls{asic}\footnote{%
	The \gls{asic} has been manufactured in a standard \SI{65}{\nano\meter} CMOS technology.
	It was conceived and designed at Heidelberg University.
	The link layer of the high-speed serial links has been developed in collaboration with the TU Dresden, who also contributed the PLL.
	The fast ADC is a result of a collaboration with the EPFL Lausanne.
}~\citep{pehle2022brainscales2_nopreprint_nourl}
is the key component of the presented system.
It is a mixed-signal implementation comprised of analog and digital building blocks (\cref{fig:system overview}) that simultaneously supports the processing of \gls{vmm} operations and the emulation of \glspl{snn} in the analog domain.
Embedded \glspl{ppu} allow for online on-chip learning.
\minisection{Analog Network Core}
\gls{bss2} contains a total of \num{512} analog neuron circuits, each receiving input from \num{256} synapses.
The neurons emulate the \gls{adex} model in \num{1000}-fold accelerated continuous time and can be combined to represent structured neurons with multiple compartments.
Each synapse contains correlation sensors enabling \gls{stdp} in \glspl{snn} and is modulated by a digital weight with \SI{6}{\bit} resolution.
For \glspl{vmm}, the neuron circuits are configured as analog accumulators, while the synapses perform multiplications.
When processing \glspl{cdnn} and \glspl{snn}, the combination of these neurons and the synapse matrix therefore perform all computations in the analog domain.
\minisection{Event Router}
The distribution of the real-time vector inputs or spike events to and from the analog network core is handled by a runtime configurable digital routing crossbar.
\minisection{Top and Bottom \glspl{ppu}}
Each chip includes two custom \SI{32}{\bit} CPUs compatible with the embedded PowerPC \gls{isa}~\citep{powerisa_206}.
They additionally feature \gls{simd} extensions for fast vector operations, which can make use of parallel ADCs (\num{1024} channels, \SI{8}{\bit} resolution) to process analog observables.
These embedded cores are primarily intended to support learning and plasticity algorithms in \glspl{snn}.
They can access most of the internal digital resources of the \gls{asic} and -- as described in \cref{sec:ecg-classification} -- serve as experiment controllers.
\minisection{Digital Core Logic}
The core control and network logic handles all off-chip communication from the embedded processors and the event router.
In addition, it bidirectionally converts between real-time and time-stamped event packets.
The transport layer manages secured memory access operations as well as unsecure, low-latency event streams over high-speed serial links to the \gls{fpga} fabric.

The right side of \cref{fig:hicann_layout} shows a layout drawing of the \gls{asic}.
The embedded processors are highlighted by the yellow rectangles.
The red frame depicts one of the four identical quadrants of the analog core.
The left side of the figure illustrates the neuromorphic processing loop through the system, together with the arrangement of neurons and synapses within a quadrant.

\begin{figure*}
	\center{
		\includegraphics[width=.9\linewidth, page=2,viewport=0 0 20.5cm 5cm, clip]{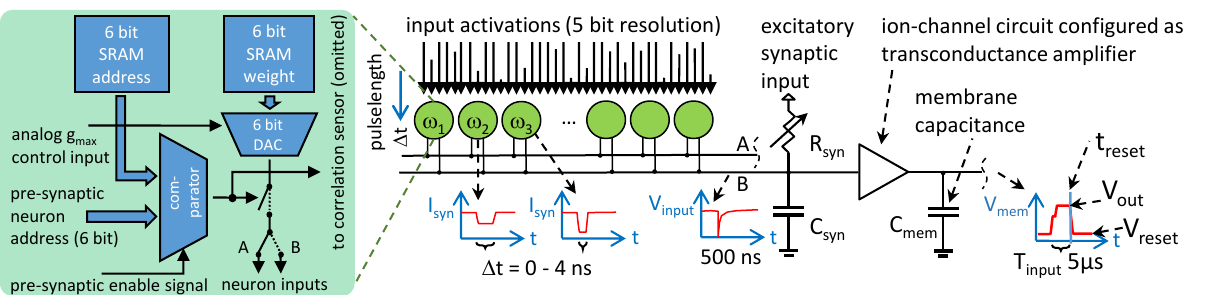}
		\caption{%
			\label{fig:synapse}
			Operation principle of \gls{cdnn} processing: the bottom half depicts the main functional blocks of a synapse circuit.
			For the \gls{vmm} calculation only the shaded area is used.
			The top half shows the analog operations taking place: each synapse generates a current pulse I\textsubscript{syn} in response to a pre-synaptic input event.
			During the calculation period T\textsubscript{input} they are integrated on the membrane capacitance.
			The final voltage V\textsubscript{out} of a single neuron represents the result of the analog \gls{vmm} calculation.
		}
	}
\end{figure*}
In \gls{cdnn} experiments, as used for the \gls{ecg} classification showcased in \cref{sec:ecg-classification}, the dataflow is as follows:
Initially, the synapse matrix is filled with weight data and the neuron circuits are configured as linear integrators without any long-term internal dynamics.
All neurons are reset to an initial membrane value $V_\text{reset}$ before the arrival of the first component of the input vector.
Inference calculation starts when the digital core logic transmits the events it has received from the \gls{fpga} to the real-time event router.
They are then distributed to synapse drivers, which in turn transmit them into the synapse array.

\Cref{fig:synapse} illustrates the principle of analog computation used for the \gls{vmm}:
To perform the analog multiplication, the events are converted from \SI{5}{\bit} binary coding to a pulse length representation.
Each synapse produces a current proportional to its \SI{6}{\bit} stored weights $\omega_x$ for the duration of the input signal they receive from the synapse drivers $\Delta t$, thereby performing an analog multiplication.
The input line of the neuron subsequently receives the sum of all output currents generated by the synapses within a vertical column.
A transconductance amplifier in each neuron generates a current equivalent to the charge received from the synapses.
Each column's current is integrated on the membrane capacitance of its associated neuron circuit.
Each neuron has two separate inputs for excitatory (A) and inhibitory (B) synaptic inputs.
For the inference calculation, they are used to represent positive and negative weight values.
For reasons of printing space, the column is shown horizontally in the figure.
For up to \num{65536} signed matrix elements, this operation is carried out in parallel within the analog core.

After an input vector has been processed in the analog domain, the neuron voltages are digitized by the parallel ADC with \SI{8}{\bit} resolution.
The \gls{relu} operation can be performed automatically during this conversion by aligning the ADC offset with the initial membrane value $V_\text{reset}$.
Alternatively, the embedded \gls{ppu} can apply an activation function to the digitized analog result, representing the output activations of a network layer.
Values that are re-used in a succeeding operation, are then converted to \SI{5}{\bit} input activations by subtracting $V_\text{reset}$ and applying bitwise right-shifts.
The results are passed to the \gls{fpga} fabric (\cref{subsec:fpga}) and either stored in \acrshort{dram} or used as inputs for the next layer.
This loop is repeatedly executed until all layers have been processed.

Each synapse can process back-to-back activations with a period of \SI{8}{\nano\second}, resulting in a maximum continuous input data rate of \SI{125}{\mega\hertz} (\cref{fig:synapse}).
There are \num{256}\texttimes{}\num{512} synapses in total, which can all simultaneously process input activations at the full data rate. This equals a maximum of
\begin{equation}
	\SI{125}{\mega\hertz}\cdot 256\cdot512\cdot \SI{2}{\op} = \SI{32.8} {\tera\op\per\second}, \label{eq:maxrate}
\end{equation}
counting multiplication and addition as individual operations.

The full integration cycle, including the necessary time to reset the neuron membrane voltages, takes about \SI{5}{\micro\second}.
This reduces the back-to-back, maximum size \gls{vmm} rate to \SI{200}{\kilo\hertz} and the resulting speed to approximately
\begin{equation}
	\frac{1}{\SI{5}{\micro\second}}\cdot 256\cdot 512\cdot\SI{2}{\op} \approx\SI{52}{\giga\op\per\second}. \label{eq:actrate}
\end{equation}

For more details on the \acrlong{bss2} architecture, we refer to~\citet{pehle2022brainscales2_nopreprint_nourl};
for the rate-based operation mode see~\citet{weis2020inference}.

\begin{figure*}
	\center{%
		\includegraphics[width=0.7\linewidth]{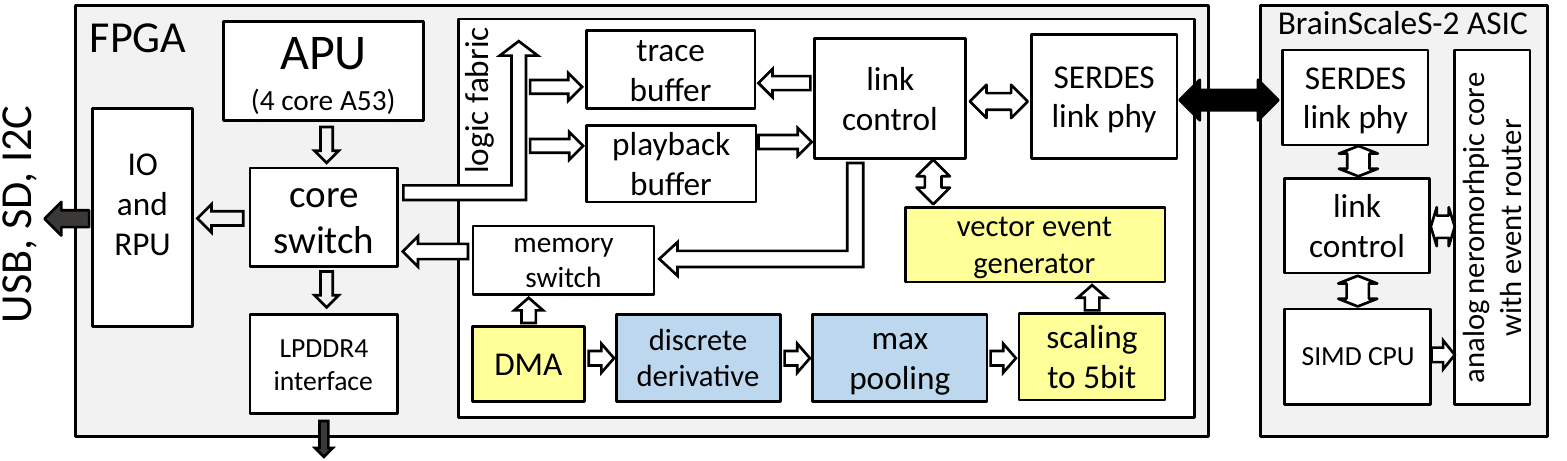}
		\caption{
			Block diagram of the major functional units of the \gls{fpga}, the part inside the logic fabric has been realized as custom \gls{rtl} in SystemVerilog.
			The \gls{dma} controller, preprocessing chain elements and vector event generator create the input activation events representing the vector in the vector-matrix multiplication.
			Some of the preprocessing (blue) is problem-specific for the medical \gls{ecg} dataset.
			To the right side the major blocks of the \gls{bss2} \gls{asic} are shown as well to illustrate the complete communication path from the embedded \glspl{ppu} to the \acrshort{dram} memory.
			The arrows denote the control flow direction from initiator to follower of the internal (hollow) and external (filled) data buses shown in the figure.
			\label{fig:fpga_logic} }}
\end{figure*}

\subsection{ASIC Adapter Board}\label{subsec:system}
The \gls{asicab} is required to interface an off-the-shelve \gls{fpga} board with the \gls{bss2} \gls{asic}.
It provides six power supply rails, three reference voltages, and a reference current to the \gls{asic}, all of which are runtime-adjustable.
The individual supply currents of the \acrlong{bss} \gls{asic} can be monitored by several shunt-based power monitoring \glspl{ic}~\citep{TexasInstruments2020INA219}.
The \gls{asic} provides eight independent bidirectional source-synchronous \gls{lvds} data channels operated at up to \SI{2}{\giga\bit\per\second} each.
Due to I/O limitations of the \gls{fpga} board, only five are routed through the \gls{asicab} to the \gls{fpga}.
Micro-SMT coaxial connectors are available for monitoring the analog outputs from the \gls{bss2} \gls{asic} as well as supplies and reference voltages.

The \gls{asic} itself is directly bonded to a carrier \gls{pcb} using a zero-insertion force \gls{sodimm} board edge connector for an optimal combination of simplicity and reliability.
\Cref{fig:system photo} shows the die bonded to the \gls{asic} carrier \gls{pcb}\@.

\subsection{System Controller}\label{subsec:fpga}
The system controller is a low-power \gls{fpga} with an embedded quad-core microprocessor~\citep{xilinx2019zynqultra} coupled with \SI{2}{\gibi\byte} of LPDDR4 \acrshort{dram}\@.
It features USB~3.0 (device \& host), SDXC, 802.11b/g/n Wi-Fi as well as Bluetooth 4.2 (BLE) communication circuits.
Further information about the \gls{fpga} base board can be found in~\citep{avnet2020ultra96}.  %

\Cref{fig:fpga_logic} depicts the internal structure of the logic fabric.
Main components are the link control and physical layer that implement the high-speed serial links to the \gls{asic}.
The playback buffer contains a list of commands to send to the \gls{asic}, while the trace buffer collects events sent back from the \gls{asic}.
Memory-mapped write and read commands can also be issued from the \gls{asic} to the \gls{fpga}.
This allows the \glspl{ppu} to access the \acrshort{dram} memory connected to the \gls{fpga} via a memory switch.

A \gls{dma} controller reads the input data from memory, converts it into input events, and sends them to the \gls{asic}.
For the experiment described in \cref{sec:ecg-classification}, this \gls{dma} controller is programmed by the \gls{ppu} on the \gls{asic} to transfer the raw signal data, an \gls{ecg} trace composed of \SI{12}{\bit} values, from memory.
The \gls{asic} requires specially formatted event data packets encoding \SI{5}{\bit} input activations for the vector-matrix multiplication.
This demands a preprocessing chain inside the \gls{fpga}, which is problem-specific to some extent.
Its function will be explained in \cref{subsec:model}.
After the raw signal data is converted into \SI{5}{\bit} values, the vector event generator attaches an event address from a lookup table.
This event is sent to the \gls{asic} via the serial links.
In the \gls{asic}, the attached addresses are used to forward the events to their target inputs of the analog neuromorphic core.
The use of a lookup table inside the \gls{fpga} allows arbitrary mapping of input vector elements onto the synapse matrix.
During the inference process the \gls{ppu} inside the \gls{asic} synchronizes the vector event generator inside the \gls{fpga} using multiple handshake signals to control the timing of the sent events.

The four \SI{64}{\bit} ARM processor cores contained in the \gls{fpga} usually do not participate in the inner loop of the inference calculation and only perform system initialization tasks.
Making use of their flexible I/O, they can however be used to form a tight, low-latency coupling between sensors, actors and the neuromorphic \acrshort{asic}.

\subsection{Software}\label{subsec:software}
Similar to other neuromorphic hardware platforms software is an essential component to make complex hardware systems accessible to users, e.g.,
\mbox{GraphCore}~\citep{kacher2020graphcore},
\mbox{Loihi}~\citep{rueckauer2021nxtf,dewolf2020nengo,lin2018programming},
\mbox{Neurogrid}~\citep{benjamin2021neurogrid,voelker2017extending},
\mbox{SpiNNaker}~\citep{rhodes2018spynnaker,rowley2019spinntools,galluppi2015framework},
\mbox{Tianjic}~\citep{ji2016neutrams},
and
\mbox{TrueNorth}~\citep{amir2013cognitive}.
A recent publication covering the older \gls{bss1} platform shortly compares software approaches of multiple neuromorphic systems~\citep{mueller2022operating}.

In each phase -- from hardware commissioning, to model design, to training, to validation -- users can take advantage of a software environment that provides appropriate abstraction levels, access to hardware debugging information as well as robust and transparent platform operation.
For the \gls{bss2} architecture, -- and, in particular, the mobile system -- we provide software support for different system aspects:

	\minisection{User Interface}
		The PyTorch toolkit~\cite{paszke2019pytorch_nourl} is a commonly used workhorse in the field.
		Particularly, it simplifies many aspects of \gls{cdnn} modeling.
		We developed a custom extension for PyTorch, \emph{hxtorch}~\cite{spilger2020hxtorch}, providing support for the \gls{bss2} architecture.
	\minisection{Training}
		Forward propagation is dispatched to the \gls{bss2} \gls{asic} while backward propagation is performed in software.
		Hence, \emph{hxtorch} enables using the \acrlong{bss2} system as an inference accelerator in PyTorch while adopting a hardware-in-the-loop-based training approach.
		The trained model can be serialized, stored to disk, and used in a \emph{standalone inference mode} to increase energy efficiency.
		In addition, a ``mock mode'' enables the simulation of certain hardware properties in software.
		This facilitates migrating from the training of a pure software model to hardware-in-the-loop-based training.
	\minisection{Hardware Resources}
		\emph{hxtorch} provides support for the execution of neural network graphs on an arbitrary number of \gls{bss2} \glspl{asic}.
		Individual layers are partitioned into chip-sized chunks and executed either in parallel, serially, or in the appropriate mixture needed to fit on the available hardware resources.
		Finally, each \gls{asic} receives and executes a stream of instructions and data.
	\minisection{Data-Flow Graph Execution}
		Internally, model layers in \emph{hxtorch} build up a data-flow graph.
		A \gls{jit} compiler traverses the graph and partitions individual layers into chunks fitting onto the available hardware resources.
		Partitioned layers are converted into configuration data and control flow statements;
		both of which are transferred to the \gls{bss2} hardware system and result data is read back.
		Regarding control flow, the hardware execution engine supports two modes:
		the first mode uses the \gls{fpga} to handle control flow;
		the second mode, which is also largely used in the standalone inference mode, hands over the control flow to the embedded \glspl{ppu} of the \gls{asic}.
	\minisection{Memory Management}
		Data input, as well as output locations, are precomputed by the \gls{bss2} software stack allowing for static memory management on the system.
		The \glspl{ppu} use the communication link to the \gls{fpga} to program the \gls{dma} engine inside the \gls{fpga} to automatically deliver the input activations from \acrshort{dram} to the analog processing cores.
		Analog operation results are read out by the processors, either held in \acrshort{sram} for temporary data, or stored back into \acrshort{dram} for output data.
	\minisection{Standalone Inference Mode} %
		The \gls{bss2} software layers are written in C++ and provide faster execution speeds compared to an interpreted high-level language such as Python.
		To create a lightweight inference flow for the energy measurements, a stand-alone version of the \emph{hxtorch} hardware graph executor was developed.
		This executor is implemented as a standalone binary and builds upon the same internal software layers and data formats as the \emph{hxtorch} extension.
		In contrast to the \gls{jit}-based execution flow, the standalone inference mode requires control flow to be handled by the embedded \glspl{ppu}.
		The processors operate on an instruction stream representing:
		data load and store operations,
		trigger operations for delivery of input activations from the \gls{fpga},
		reading out the neuron membrane values,
		or performing digital operations that are not supported by the analog substrate.
	\minisection{Embedded System Environment}
		The \gls{bss2} mobile system includes a Linux environment\footnote{Petalinux; the build flow of the embedded Linux distribution is provided by the \gls{fpga} manufacturer.} running on an embedded ARM64 processor.
		We take advantage of a fully containerized software environment based on singularity~\citep{kurtzer2017singularity} and spack~\cite{gamblin2015spack} to provide a cross-compiler environment on the host computer as well as on the embedded Linux system.
		Standard Linux drivers (xHCI, mass storage, FAT32) are used to read out test data from a USB mass storage device;
		additionally, support for USB-based Ethernet networking hardware is enabled to facilitate remote system usage.
		An experiment execution service enables users to run Python-based interfaces on host computers that exchange serialized experiment configurations and result data with the mobile system.

Details on \emph{hxtorch} for rate-based hardware operation can be found in~\citet{spilger2020hxtorch}.
A general overview of the software stack for \gls{bss2}, including spiking hardware operation, can be found in~\citet{mueller2022scalable}.

	\section{Showcase: ECG classification}\label{sec:ecg-classification}

We showcase the \gls{bss2} mobile system by classifying \gls{afib} in the medical \gls{ecg} dataset introduced in \cref{dataset}.
This real-world task demonstrates many of the platform's features, such as stand-alone operation, mobility, power efficiency and external connectivity.
We deploy a trained model on the system, which then autonomously classifies \gls{ecg} data supplied via a USB connection.

\subsection{Model}\label{subsec:model}
The model design is mainly governed by network size trade-offs between high accuracy and short runtime.
Networks that exceed the size of the compute substrate pose a high runtime and I/O penalty due to frequent reconfiguration.
This issue especially becomes relevant for non-batched operation, while it diminishes for large batch sizes.
Targeting edge applications, we restrict the inference runs to a batch size of one.

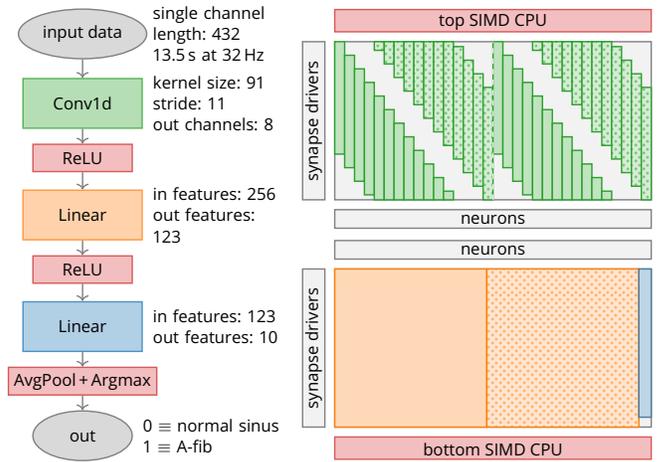
\begin{figure}[ht]
    \resizebox{\linewidth}{!}{\begin{tikzpicture}
\sffamily
\tikzstyle{layernode} = [
    rectangle,
    minimum width=2.4cm,
    minimum height=1cm,
    text centered,
    thick,
    node distance=1.15cm
]
\tikzstyle{convnode} = [layernode]
\tikzstyle{linearnode} = [layernode]
\tikzstyle{ppunode} = [
    layernode,
    minimum width=2cm,
    minimum height=.55cm,
    draw=color3!90!gray!80,
    fill=color3!30,
    node distance=1.1cm,
]
\tikzstyle{startnode} = [
    layernode,
    ellipse,
    minimum width=1.9cm,
    draw=gray,
    fill=lightgray!60,
]
\tikzstyle{endnode} = [
    layernode,
    ellipse,
    minimum width=2cm,
    minimum height=1cm,
    draw=gray,
    fill=lightgray!60,
    node distance=1.1cm,
]

\node (in) [startnode] {input data};
\node (conv1d) [convnode,node distance=1.4cm,draw=color2!80,fill=color2!35,below of=in] {Conv1d};
\node (relu) [ppunode, below of=conv1d] {ReLU};
\node (linear1) [linearnode,draw=color1!80,fill=color1!35,below of=relu] {Linear};
\node (relu1) [ppunode, below of=linear1] {ReLU};
\node (linear2) [linearnode,draw=color0!80,fill=color0!35,below of=relu1] {Linear};
\node (argmax) [ppunode, below of=linear2] {AvgPool\,+\,Argmax};
\node (out) [endnode, below of=argmax] {out};

\begin{scope}[->,thick,gray]
\draw (in) -- (conv1d);
\draw (conv1d) -- (relu);
\draw (relu) -- (linear1);
\draw (linear1) -- (relu1);
\draw (relu1) -- (linear2);
\draw (linear2) -- (argmax);
\draw (argmax) -- (out);
\end{scope}

\node at (in.center) [anchor=west,text width=2.7cm,align=left,xshift=13mm] {
    single channel\\
    length: 432\\
    13.5\,s at 32\,Hz\\
};
\node at (conv1d.center) [anchor=west,text width=2.7cm,align=left,xshift=13mm] {
    kernel size: 91\\
    stride: 11\\
    out channels: 8\\
};
\node at (linear1.center) [anchor=west,text width=2.7cm,align=left,xshift=13mm] {
    in features: 256\\
    out features: 123\\
};
\node at (linear2.center) [anchor=west,text width=2.7cm,align=left,xshift=13mm] {
    in features: 123\\
    out features: 10\\
};
\node at (out.center) [anchor=west,text width=3.2cm,align=left,xshift=11mm] {
    0 $\equiv$ normal sinus\\
    1 $\equiv$ \gls{afib}\\
};
\end{tikzpicture}

\hspace{-0.4cm}
\begin{tikzpicture}[every node/.style={inner xsep=0,outer sep=0}, scale=0.0249]
\sffamily
\def\kernelsize{91}
\def\channels{8}
\def\stride{11}
\def\expansions{32}
\pgfmathtruncatemacro{\turns}{ceil((\kernelsize + \stride * (\expansions - 1)) / 128)}

\def\patterni{none}
\def\patternii{crosshatch dots}

\draw[thick,color=gray,fill=gray!10] (-26,28) rectangle ++(18,128);
\node [rotate=90] at (-17,92) {synapse drivers};

\draw[thick,color=gray,fill=gray!10] (0,28) rectangle ++(256,128);

\begin{scope}[thick,color=color2,pattern color=color2!60]
\def\fillcolori{color2!30}
\def\fillcolorii{color2!20}
\foreach \i in {0,...,3} {
    \draw[pattern=\patterni,preaction={fill=\fillcolori}] (\i * \channels, 28 + 128 - \i * \stride) rectangle ++(\channels,-\kernelsize);
}
\foreach \i in {4,...,11} {
    \draw[pattern=\patterni,preaction={fill=\fillcolori}] (\i * \channels, 28 + 128 - \i * \stride) rectangle (\i * \channels + \channels, 28);
    \draw[pattern=\patternii,preaction={fill=\fillcolorii}] (\i * \channels, 28 + 128) rectangle (\i * \channels + \channels, 28 + 256 - \i * \stride - \kernelsize);
}
\foreach \i in {12,...,15} {
    \draw[pattern=\patternii,preaction={fill=\fillcolorii}] (\i * \channels, 28 + 256 - \i * \stride) rectangle ++(\channels,-\kernelsize);
}
\end{scope}

\begin{scope}[thick,color=color2,pattern color=color2!60]
\def\fillcolori{color2!30}
\def\fillcolorii{color2!20}
\foreach \i in {0,...,3} {
    \draw[pattern=\patterni,preaction={fill=\fillcolori}] (128 + \i * \channels, 28 + 128 - \i * \stride) rectangle ++(\channels,-\kernelsize);
}
\foreach \i in {4,...,11} {
    \draw[pattern=\patterni,preaction={fill=\fillcolori}] (128 + \i * \channels, 28 + 128 - \i * \stride) rectangle (128 + \i * \channels + \channels, 28);
    \draw[pattern=\patternii,preaction={fill=\fillcolorii}] (128 + \i * \channels, 28 + 128) rectangle (128 + \i * \channels + \channels, 28 + 256 - \i * \stride - \kernelsize);
}
\foreach \i in {12,...,15} {
    \draw[pattern=\patternii,preaction={fill=\fillcolorii}] (128 + \i * \channels, 28 + 256 - \i * \stride) rectangle ++(\channels,-\kernelsize);
}
\end{scope}
\draw[dashed,thick,color=white!90!black] (128, 28) -- ++(0, 128);

\draw[thick,color=gray,fill=gray!10] (0,5) rectangle ++(256,15);
\node at (128,12.5) {neurons};

\draw[thick,color=gray,fill=gray!10] (-26,-28) rectangle ++(18,-128);
\node [rotate=90] at (-17,-92) {synapse drivers};

\draw[thick,color=gray,fill=gray!10] (0,-28) rectangle ++(256,-128);

\begin{scope}[thick,color=color1,pattern color=color1!60]
\def\fillcolori{color1!30}
\def\fillcolorii{color1!20}
\draw[pattern=\patterni,preaction={fill=\fillcolori}] (0,-28) rectangle ++(123,-256/2);
\draw[pattern=\patternii,preaction={fill=\fillcolorii}] (123,-28) rectangle ++(123,-256/2);
\end{scope}

\draw[thick,color=color0!50!gray,fill=color0!30] (123+123,-28) rectangle ++(10,-120);

\draw[thick,color=gray,fill=gray!10] (0,-5) rectangle ++(256,-15);
\node at (128,-12.5) {neurons};

\draw[thick,color=color3!90!gray!80,fill=color3!30] (0,+8+128+28) rectangle ++(256,18);
\node at (128,+8+128+28+9-1) {top \gls{ppu}};

\draw[thick,color=color3!90!gray!80,fill=color3!30] (0,-8-128-28) rectangle ++(256,-18);
\node at (128,-8-128-28-9-1) {bottom \gls{ppu}};

\end{tikzpicture}}
    \caption{\label{fig:model}%
    Layer structure (left) and on-chip arrangement (right) of the used deep convolutional neural network model.
    The convolutional layer (green) is processed in the upper synapse array, the identical weight is arranged \num{32} times on the substrate to enable parallel processing.
    All \glspl{relu} (red) are performed in digital logic by the two \glspl{ppu}.
    The further processing takes place on the lower synapse array with a fully connected layer and \num{123} hidden neurons (orange).
    To ensure efficient use of the substrate, it is divided into two parts and placed side by side.
    The dotted part of the layer receives the second half of inputs at the same time and is processed in parallel.
    The actual classification is then achieved in the last layer (blue) with \num{10} neurons on the right, which are combined into two logical neurons by average pooling, effectively reducing analog noise.
    }
\end{figure}
Evaluation of network models showed that a small network that fits on a single chip and does not require reconfiguration can achieve reasonable classification performance.
The network used in this showcase is depicted in \cref{fig:model}.
It operates on \SI{13.5}{\second} of the \SI{120}{\second} long \gls{ecg} records, as this has turned out to be sufficient for classification of \gls{afib}.
To the left, the graph of the model is shown.
It consists of one convolutional and two linear layers.
The small size of the network allows it to be completely realized on the \gls{asic}.
The calculations in its convolutional first layer can be performed fully in parallel, as well as those in the second and third layers:
this mapping to the two halves of a \gls{bss2} \gls{asic} is shown on the right side of the figure.
The \gls{relu} and the final argmax operations are performed in the embedded \glspl{ppu} after digital readout of the analog neuron membrane voltages (cf.~\cref{subsec:neuromorphic-asic})\@.

\begin{figure}
    \center{%
        \resizebox{\linewidth}{!}{\sansmath\input{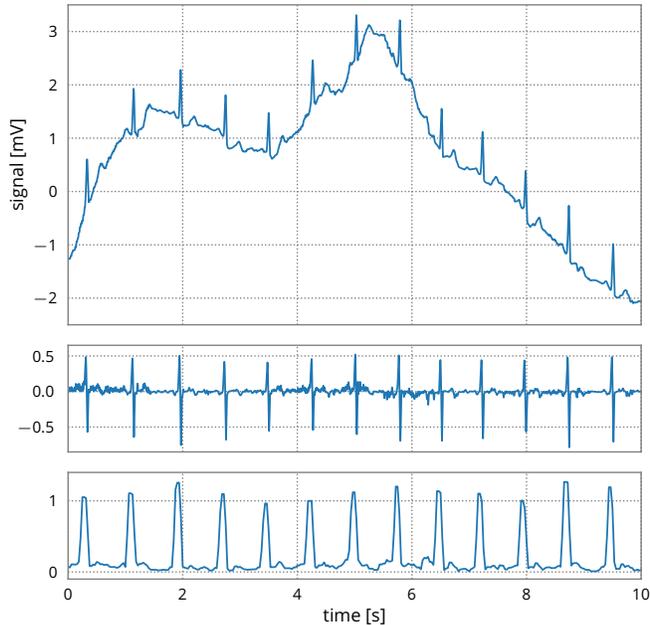}}
        \caption{Preprocessing steps performed in the \gls{fpga} fabric (from top to bottom): The raw, i.e.\ unprocessed, input sample is transformed by taking a discrete derivative to reduce baseline fluctuations. Subsequent maximum-minimum difference pooling reduces the sample rate and provides positive activations, which form the final input signal to the \gls{cdnn} in the \gls{asic}.
        Original data taken from~\citet{clifford2017af}.
        \label{fig:preprocessing} }}
\end{figure}
The \gls{asic} operates on positive activations with \SI{5}{\bit} resolution.
Since the raw data samples as input for the inference calculation are provided as \SI{12}{\bit} values with higher dynamic range, some preprocessing is required.
\Cref{fig:preprocessing} illustrates the performed steps.
To avoid unnecessary data movement, the preprocessing is done in the \gls{fpga} fabric by a custom processing chain. %
In the first step of the preprocessing, a discrete derivative of the original signal is calculated to suppress the large baseline fluctuations of the signal.
In a second step, the data rate is reduced by calculating the difference between the maximum and the minimum of \num{32} samples.
The resulting samples are quantized to \SI{5}{\bit} and used as inputs to the analog vector-matrix multiplications performed within the \gls{asic}.

\subsection{Training}\label{subsec:training}
Training relies on the proven backpropagation algorithm for \glspl{cdnn}~\citep{rumelhart86backprop}. %
To facilitate fast prototyping when training the network described in \cref{subsec:model}, a mathematical abstraction of the hardware operations was implemented on top of PyTorch~\cite{paszke2019pytorch_nourl} in \emph{hxtorch}~\citep{spilger2020hxtorch}.
Incorporating hardware-related constraints like fixed-pattern noise and limited dynamic range, it enables the training of initial models in software and provides gradient information for the backward-pass when training on hardware.
Final model parameters as presented in \cref{sec:results}, however, were trained on the \gls{asic} following a hardware-in-the-loop approach~\citep{schmitt2017hwitl_nourl}:
The forward pass is evaluated on \gls{bss2}, whereas the backward pass and parameter updates are calculated on the host computer using \emph{hxtorch}.
Tensor data structures are seamlessly converted to hardware resolution and back.
Data partitioning and experiment control is handled by both on-chip \glspl{ppu} (see \cref{subsec:software}).
To the user, the training procedure is completely embedded within PyTorch.
To increase robustness and decrease sensitivity to hardware variations, we replace the average pooling in the last layer by a max pooling operation during training.
We employ early stopping whenever no substantial improvement is observed between training epochs.

	\section{Results}\label{sec:results}

The performance of the presented system has been evaluated by assigning a set of \gls{ecg} traces to two classes: patients with sinus rhythm and patients showing \acrlong{afib} (\cref{dataset}).
Mimicking the expected workload in a low-energy edge application, all data has been processed with a batch size of one.
To increase the accuracy of all measurements, data was processed in blocks of \num{500} traces.
For each block, runtime and energy consumption have been measured using the sensors described in \cref{subsec:system} and afterwards averaged down to a single inference.
The power consumption was measured with a sampling rate of \SI{294}{\hertz} for sensors on the system controller and \SI{4.4}{\kilo\hertz} for sensors on the \gls{asicab}.

\begin{figure}
    \center{%
        \resizebox{\linewidth}{!}{\sansmath\input{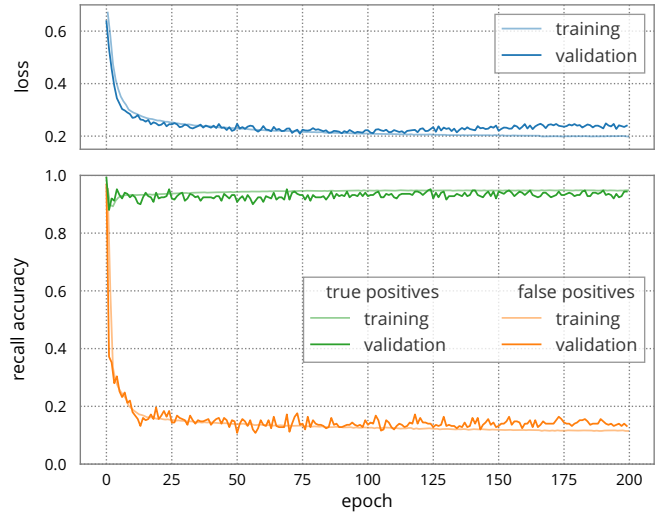}}
        \caption{\label{fig:ecg_training_statistics}%
            Training and validation metrics of the model presented in \cref{fig:model} performed with the \gls{bss2} \gls{asic}.
            The test set of 500 records was split from the provided \gls{ecg} dataset prior to training.}}
\end{figure}

Classification accuracy has been evaluated by selecting randomized test sets of \num{500} records prior to training.
Metrics of such a training course on the presented system are shown in \cref{fig:ecg_training_statistics}.
With the shown combination of model, software and hardware, this system classified \gls{afib} with a detection rate of \SI{93.7(7)}{\percent} at \SI{14.0(10)}{\percent} false positives.

Each block of \num{500} input traces was found to be processed in \SI{138}{\milli\second};
starting with raw \gls{ecg} data in the system controller \acrshort{dram} and ending with binary classification results ibidem.
\Cref{tab:competition_results} gives an overview over the achieved results:
During the inference phase, the system achieved \SI{477}{\mega\op\per\second} with a mean power consumption of \SI{5.6}{\watt}.
In its current state, classification on the \gls{bss2} mobile system takes \SI{276}{\micro\second} and consumes a total of \SI{1.56}{\milli\joule} per \gls{ecg} trace, of which \SI{192}{\micro\joule} were consumed by the \gls{bss2} \gls{asic}.

\begin{table}
	\center{
		\caption{%
			Measured results for the classification of a single \gls{ecg} trace.
			To increase measurement precision, data has been acquired as a block of \num{500} traces, which were classified in direct succession with batch size one.
			Unless noted otherwise, the given number represents the mean value from this set.
			The tested records have been excluded from training.
			\label{tab:competition_results}
		}
		\begin{tabular}[b]{lr@{\hspace{5pt}}ll}
			\toprule
			quantity & \multicolumn{2}{r}{value} & unit \\
			\midrule
			time per inference                                       & \num{276}  & \num{e-6} & \si{\second}        \\
			power consumption (system)                               & \num{5.6}  &           & \si{\watt}          \\
			power consumption  (\gls{bss2} \gls{asic})               & \num{0.69} &           & \si{\watt}          \\[1mm]
			energy (total)                                           & \num{1.56} & \num{e-3} & \si{\joule}         \\  %
			$~~$energy (system controller, total)                    & \num{0.7}  & \num{e-3} & \si{\joule}         \\  %
			$~~~~$energy (system controller, ARM \acrshort{cpu})     & \num{0.34} & \num{e-3} & \si{\joule}         \\  %
			$~~~~$energy (system controller, \gls{fpga})             & \num{0.21} & \num{e-3} & \si{\joule}         \\  %
			$~~~~$energy (system controller, \acrshort{dram})        & \num{0.12} & \num{e-3} & \si{\joule}         \\  %
			$~~$energy (\gls{asic}, total)                           & \num{0.19} & \num{e-3} & \si{\joule}         \\  %
			$~~~~$energy (\gls{asic}, IO)                            & \num{0.07} & \num{e-3} & \si{\joule}         \\  %
			$~~~~$energy (\gls{asic}, analog)                        & \num{0.07} & \num{e-3} & \si{\joule}         \\  %
			$~~~~$energy (\gls{asic}, digital)                       & \num{0.07} & \num{e-3} & \si{\joule}         \\[1mm]  %
			total operations in \gls{cdnn}                           & \num{132}  & \num{e3}  & \si{\op}            \\
			\gls{bss2} \gls{asic} processing speed (mult./acc.)      & \num{477}  & \num{e6}  & \si{\op\per\second} \\
			\gls{bss2} \gls{asic} energy efficiency (mult./acc.)     & \num{689}  & \num{e6}  & \si{\op\per\joule}  \\
			\gls{bss2} \gls{asic} energy efficiency (inferences)     & \num{5.25} & \num{e3}  & \si{\per\joule}     \\[1mm] %
			classification accuracy                                  &            &           &                     \\
			$~~$detection rate & \multicolumn{2}{c}{\num{93.7(7)}}   & \si{\percent} \\
			$~~$false positives & \multicolumn{2}{c}{\num{14.0(10)}} & \si{\percent} \\
			\bottomrule
		\end{tabular}
	}
\end{table}

	\section{Discussion \& Conclusions}\label{sec:conclusion}
We have presented the \gls{bss2} mobile system as an analog inference platform and demonstrated medical \gls{ecg} data classification as one possible application.

The small system is mobile by design and has proven to operate reliably under various environmental conditions.
Despite its early prototype stage, it is therefore directly applicable to inference tasks on the edge:
The results we have achieved demonstrate that the presented system is sufficiently energy-efficient to run on battery while monitoring the health of a patient.
Based on the energy consumption presented in \cref{tab:competition_results}, a common CR2032 lithium button battery with an approximated energy content of \SI{200}{\milli\ampere\hour} would power the inference calculations for detecting atrial fibrillation in two-minute intervals for five years.
At the cost of runtime and thus energy efficiency, we can utilize larger networks to increase the classification accuracy.
On the \gls{bss2} \gls{asic}, we have achieved accuracies of up to \SI{95.5}{\percent} for \gls{afib} with \SI{8.0}{\percent} false positives.

The achieved detection rates on the \gls{bss2} mobile system are on par with other state-of-the-art solutions:
\citet{rizwan2021review} report \acrlong{afib} detection rates for machine-learning-based solvers from \SIrange{80.0}{100.0}{\percent} with a median of \SI{96.3}{\percent} (\SIrange{1.09}{26.4}{\percent} false positives, median: \SI{6.9}{\percent}).
Solvers based on classical time series analysis reach \SIrange{74.2}{99.6}{\percent} with a median of \SI{97.1}{\percent} (\SIrange{1.7}{10.2}{\percent} false positives, median: \SI{3.2}{\percent}), as presented by~\citet{marsili2020implementation}.
Most of these solutions, however, do not target the low power envelope required for edge applications.
In contrast, \citet{azariadi2016ecg,seitanidis2022identifying} use the off-the-shelve Intel Galileo and Nvidia Jetson Nano platforms to classify \gls{ecg} anomalies with an energy consumption of \SI{220}{\milli\joule} and \SI{7.4}{\milli\joule} per inference.\footnote{
    We assume a power consumption of \SI{2.2}{\watt} for the Intel Galileo and \SI{5.0}{\watt} for the Nvidia Jetson Nano system and use the published inference runtimes to estimate the energy per inference.
}
With a similar system controller and power consumption, the presented \gls{bss2} mobile system only consumes \SI{1.56}{\milli\joule} per classification.
Designed as a generic computational substrate for a multitude of applications, it can however not compete with \glspl{asic} specifically built for low-power \gls{afib} classification:
\citet{andersson2015subthreshold} present a classifier that achieves a comparable detection rate of \SI{94.9}{\percent} (\SI{4.7}{\percent} false positives) with a power envelope of only \SI{334}{\nano\watt}.

In addition to the presented multiply-accumulate functionality, \gls{bss2} is designed to operate as an analog emulator for \glspl{snn}.
\citet{cramer2022surrogate} present classifiers on multiple common datasets that make use of this mode to achieve beyond state-of-the-art classification latency and energy efficiency on \gls{bss2}.
To the best of our knowledge, it is the first and only available system to accelerate both, multiply-accumulate operations and \glspl{snn} in the analog domain.
Due to the stateful nature of the necessary time-continuous operations, multiplexing of analog resources is seldom possible in \gls{snn} accelerators, therefore limiting the maximum model size to the available hardware resources.
In contrast, rate-based stateless operation using our analog neuromorphic core as a parallel vector-matrix multiplier allows for multiplexing hardware resources in time and therefore has the advantage of supporting arbitrarily large model sizes.
Such networks are only limited by the available memory.
Most models that are capable of performing real-world tasks, like video analysis or speech translation, need model sizes in the order of \numrange{e7}{e9} parameters~\citep{aharoni2019massively}.
These network sizes are feasible with the presented system, as neither the hardware platform nor the \emph{hxtorch} software environment impose size limitations on the model in use.

The combination of spiking and convolutional neural networks on a single substrate therefore greatly widens the application of \glspl{snn} in edge applications:
it allows features to be extracted by conventional high dimensional \gls{cdnn} layers on multiplexed hardware resources, while sparse spiking layers can simultaneously be used for their final classification.
Using the embedded \glspl{ppu}, \gls{bss2} can utilize online learning for the \gls{snn} layers~\citep{pehle2022brainscales2_nopreprint_nourl} and thereby improve classification performance and adapt to environmental changes in the field.

Given its early prototyping stage, the system as well as the \gls{bss2} chip itself contain a large potential for optimization.
Currently, the \gls{fpga} is primarily used as a memory controller for the \gls{asic} -- functionality that could be incorporated into the chip's digital core.
This would remove the power consumption of the \gls{fpga} from the system's energy balance and would increase the bandwidth between memory and analog core. %

The main motivation during the development of the \gls{bss2} \gls{asic} was to enable flexible on-chip online learning in \glspl{snn}.
Thus, the speed of the analog \gls{cdnn} calculation has not yet been optimized.
While the synapse arrays that perform the multiply-accumulate operation already support \SI{32.8}{\tera\op\per\second}, see~\cref{eq:maxrate}, the usage of the spike-based neurons for the integration of the summation currents limits the actual speed to approximately \SI{52}{\giga\op\per\second}, see~\cref{eq:actrate}.

The current area efficiency of the analog MAC in the synapse arrays can be calculated as
\begin{equation}
    \frac{\SI{32.8}{\tera\op\per\second}}
    {256\cdot 512\cdot \SI{8}{\micro\meter}\cdot \SI{12}{\micro\meter}} = \SI{2.6}{\tera\op\per\second\per\square\milli\meter}.
\end{equation}
As a conservative approximation based on the current die size of \SI{32}{\square\milli\meter}, we target an area efficiency above \SI{1}{\tera\op\per\second\per\square\milli\meter} for the full chip.
State-of-the-art implementations using similar technologies and architectures reach up to \SI{0.32}{\tera\op\per\second\per\square\milli\meter} based on full die size~\citep{fick2022analog,kim2022moneta}.

Multiple approaches have to be taken to make use of the aforementioned processing speed of the synapse array:
First, specialized circuits for the integration of the synapses' output currents in the non-spiking operation mode of the \gls{asic} have to be integrated.
These specialized accumulators could be combined with revised parallel ADCs that are -- in contrast to the currently implemented design -- capable of sufficient conversion speed.
The increased data rate will require higher I/O bandwidth that could be achieved by the aforementioned integration of an on-chip memory controller.

\label{availability-and-access}
In its current state, the \acrlong{bss2} system is available to the scientific community via the EBRAINS project\footnote{\url{https://ebrains.eu/register}}.
Example applications using \glspl{snn} as well as the built-in multiply-accumulate functionality are available and can be executed online through a browser-based interface.
Hardware access to the \gls{bss2} (mobile) system is available upon request.

	\section*{Contributions}%
\label{sec:contributions}

Yannik Stradmann directed the development and modeling efforts for the presented experiment and hardware setup.
He contributed to all components.
Sebastian Billaudelle contributed to the chip design, chip commissioning and implementation of the experiment.
Oliver Breitwieser contributed to the software stack, is the main architect of the preemptive experiment scheduling service and contributed to modeling and model verification.
Falk Ebert is a main contributor to the energy measurement system.
Arne Emmel developed and implemented the model, designed the preprocessing, adapted the training to the hardware platform and contributed to the software integration.
Dan Husmann developed the \gls{asicab}\@.
Joscha Ilmberger is the main system developer contributing to \gls{pcb} design, porting of the \gls{fpga} design to the new platform and adding functionality such as preprocessing and the vector event generator.
Eric Müller is the lead developer and architect of the \gls{bss2} software stack;
he commissioned the embedded platform, ported the software development environment as well as the \gls{bss2} software stack to the embedded \gls{fpga} platform.
Philipp Spilger is the main developer of the software for the non-spiking operation mode of the \gls{bss2} \gls{asic} and a contributor to the software stack.
Johannes Weis is the main developer of calibration routines for the analog network core, commissioned the first non-spiking experiments on the hardware platform and contributed to the model.
Johannes Schemmel is the lead designer and architect of the \gls{bss2} neuromorphic system. He wrote the initial version of the paper.
All authors contributed to and edited the final manuscript.

\section*{Acknowledgments}
\label{sec:acknowledgements}
The authors wish to thank all present and former members of the Electronic Vision(s) research group contributing to the BrainScaleS-2 neuromorphic platform.
For the publication fee the authors acknowledge financial support by the \foreignlanguage{ngerman}{Deutsche Forschungsgemeinschaft} (DFG, German Research Foundation) within the funding programme \foreignlanguage{ngerman}{\glqq{}Open Access Publikationskosten\grqq{}} as well as by Heidelberg University.

	{
	\footnotesize%
	\bibliography{main}
	}

	\begin{IEEEbiography}[{\includegraphics[width=1in,height=1.25in,clip,keepaspectratio]{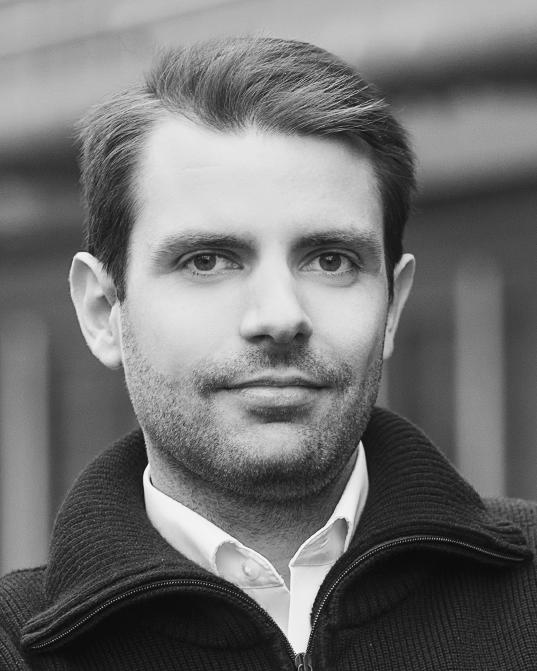}}]{Yannik Stradmann }
	received the M.Sc.\ degree in Physics from Heidelberg University, Germany, in 2019.
	Currently, he is a Ph.D.\ student in the Electronic Vision(s) group at Heidelberg University.
	His research focuses on the development and characterization of mixed-signal VLSI circuits for neuromorphic hardware and their application for real-time control tasks.
\end{IEEEbiography}

\begin{IEEEbiography}[{\includegraphics[width=1in,height=1.25in,clip,keepaspectratio]{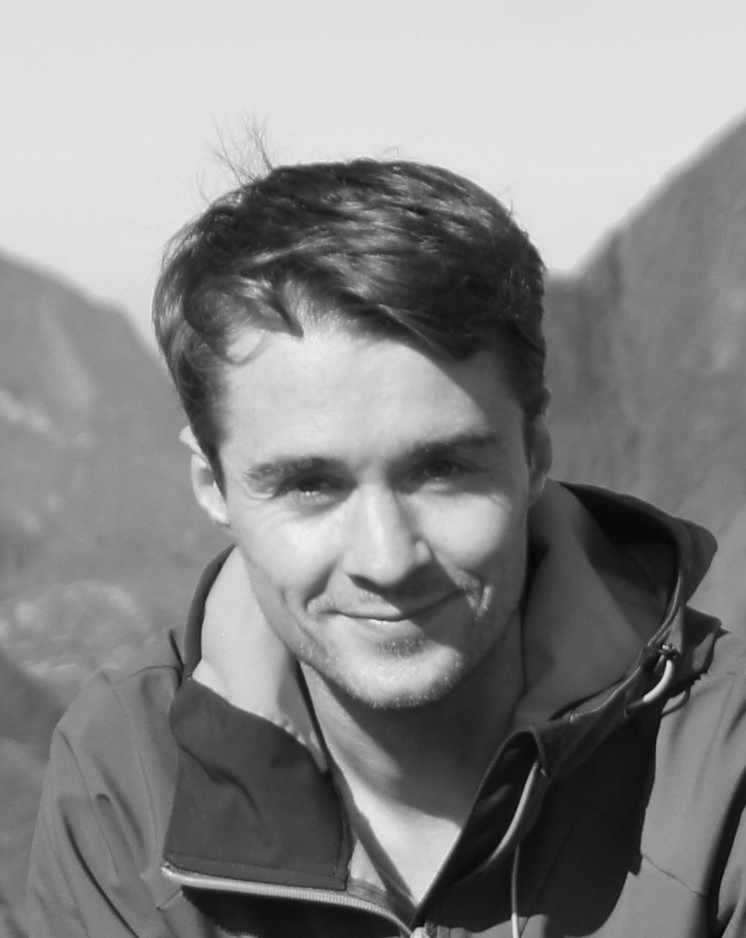}}]{Sebastian Billaudelle }
	received his Ph.D.\ degree in physics from Heidelberg University, Germany in 2022.
	As a postdoc in the Electronic Vision(s) group at Heidelberg University, he contributes to the analog neuromorphic circuits of BrainScaleS-2 and devises training and learning algorithms for the neuromorphic system.
\end{IEEEbiography}

\begin{IEEEbiography}[{\includegraphics[width=1in,height=1.25in,clip,keepaspectratio]{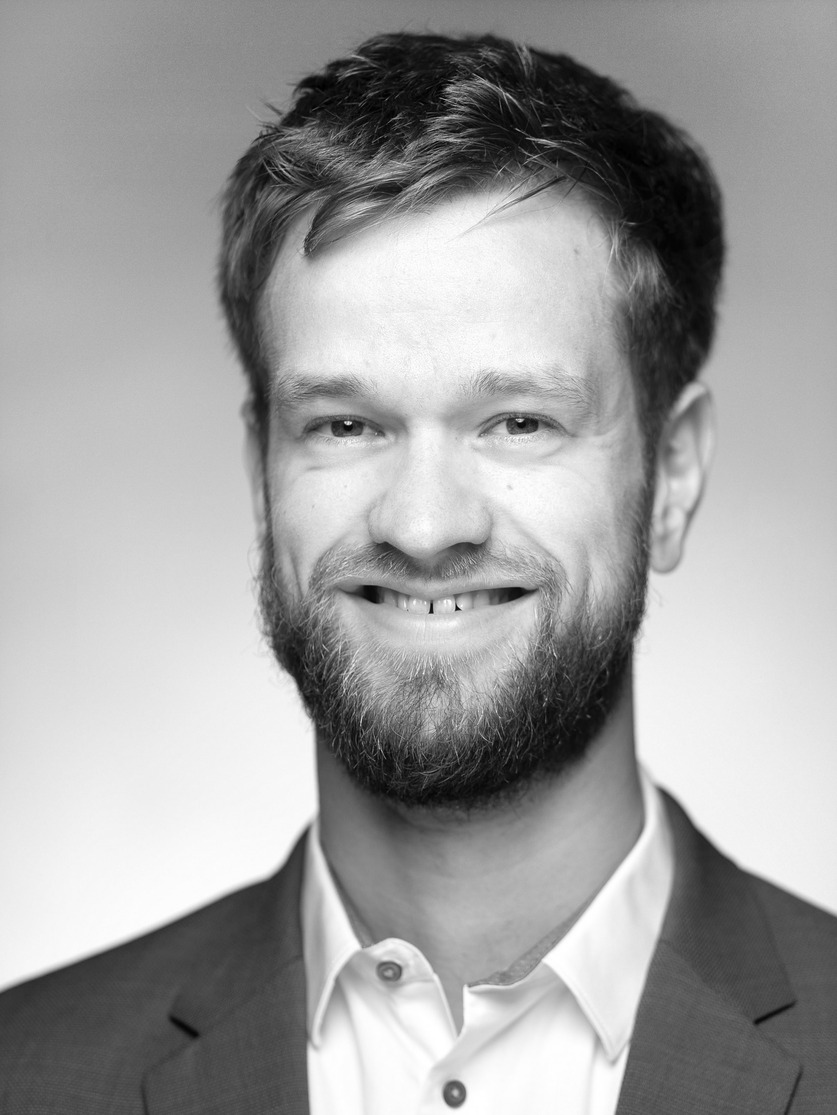}}]{Oliver Breitwieser }
  received the Ph.D.\ degree in Physics from Heidelberg University, Germany, in 2021.
  His research interests include neuroscience, machine learning, particularly applied to neuromorphic hardware, as well as sustainable distributed large-scale computing.
  He is a core developer of the BrainScaleS operating system and responsible for operations of the BrainScaleS compute infrastructure.
\end{IEEEbiography}

\begin{IEEEbiography}[{\includegraphics[width=1in,height=1.25in,clip,keepaspectratio]{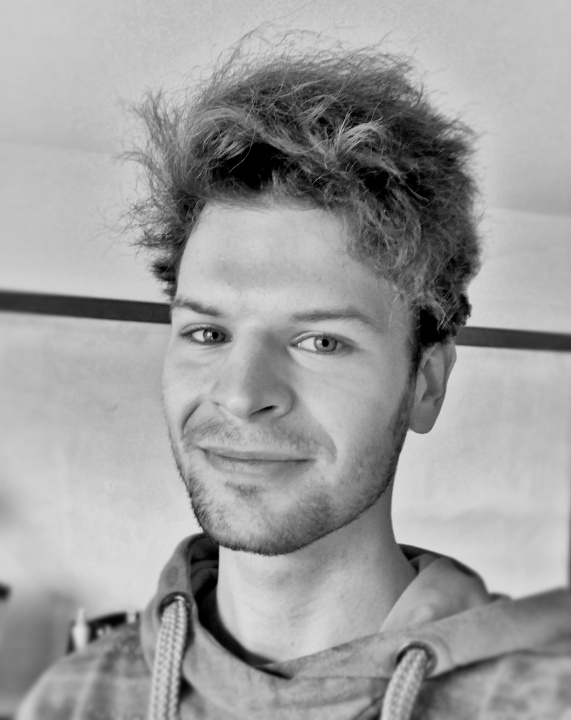}}]{Falk Leonard Ebert }
	is studying physics at Heidelberg University, Germany and currently working on his B.Sc.\ thesis in the Electronic Vision(s) group at the
	Kirchhoff-Institute for Physics.
\end{IEEEbiography}

\begin{IEEEbiography}[{\includegraphics[width=1in,height=1.25in,clip,keepaspectratio]{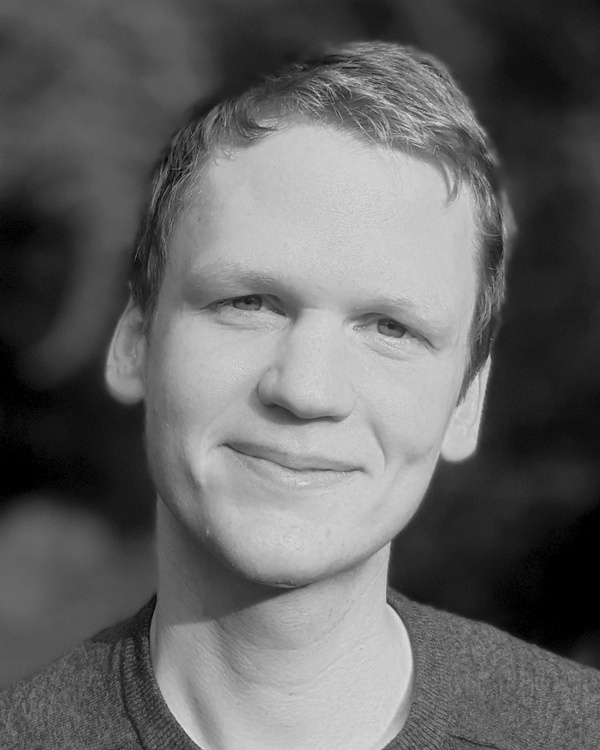}}]{Arne Emmel }
	received the M.Sc.\ degree in physics from Heidelberg University, Germany, in 2020.
	Currently, he is a researcher in the Electronic Vision(s) group at Kirchhoff-Institute for Physics, Heidelberg University.
	His interests include modeling and the exploration and optimization of training algorithms for the specific requirements of analog neuromorphic hardware.
\end{IEEEbiography}

\begin{IEEEbiography}[{\includegraphics[width=1in,height=1.25in,clip,keepaspectratio]{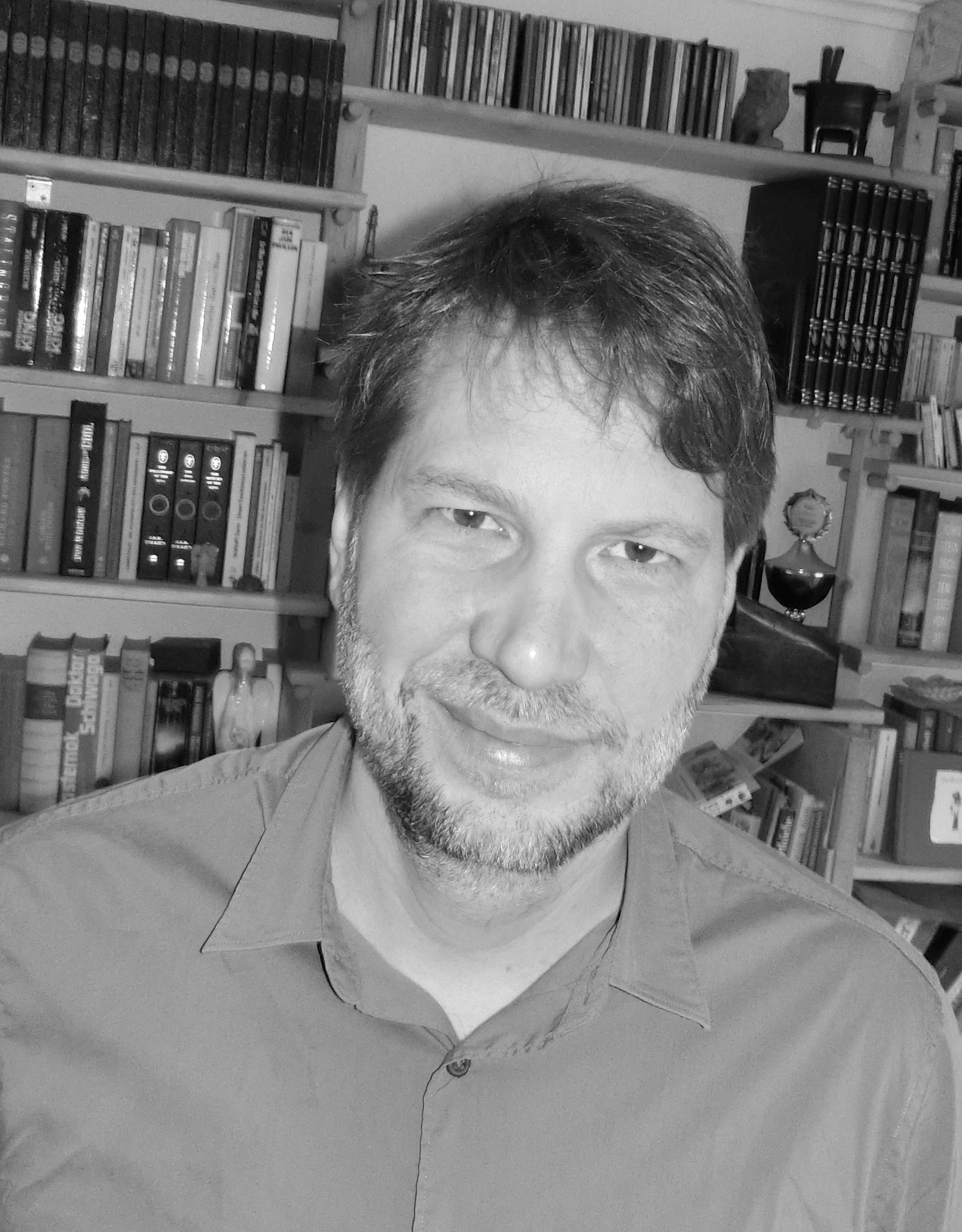}}]{Dan Husmann }
    received the diploma\ degree in physics from Heidelberg University, Germany, in 2000.
	Currently, he is a researcher in the Electronic Vision(s) group at Kirchhoff-Institute for Physics, Heidelberg University.
    His research interest are wafer-scale integration techniques and building of neuromorphic hardware.
\end{IEEEbiography}

\begin{IEEEbiography}[{\includegraphics[width=1in,height=1.25in,clip,keepaspectratio]{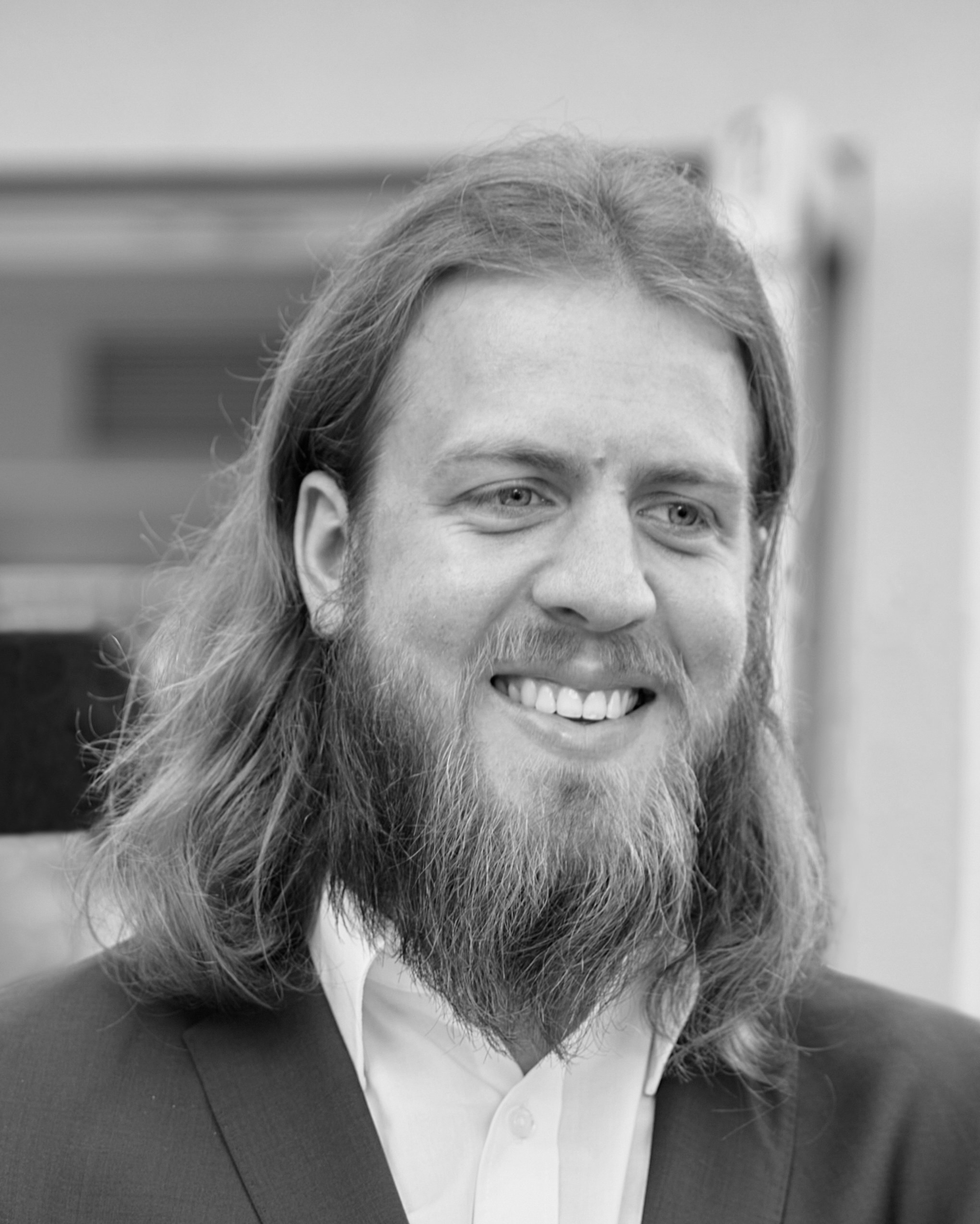}}]{Joscha Ilmberger }
	received his M.Sc.\ degree in physics from Heidelberg University, Germany in 2017.
	As a Ph.D.\ student in the Electronic Vision(s) group at Heidelberg University, his research interests are scaling and digital architecture design of novel analog neuromorphic hardware.
\end{IEEEbiography}

\begin{IEEEbiography}[{\includegraphics[width=1in,height=1.25in,clip,keepaspectratio]{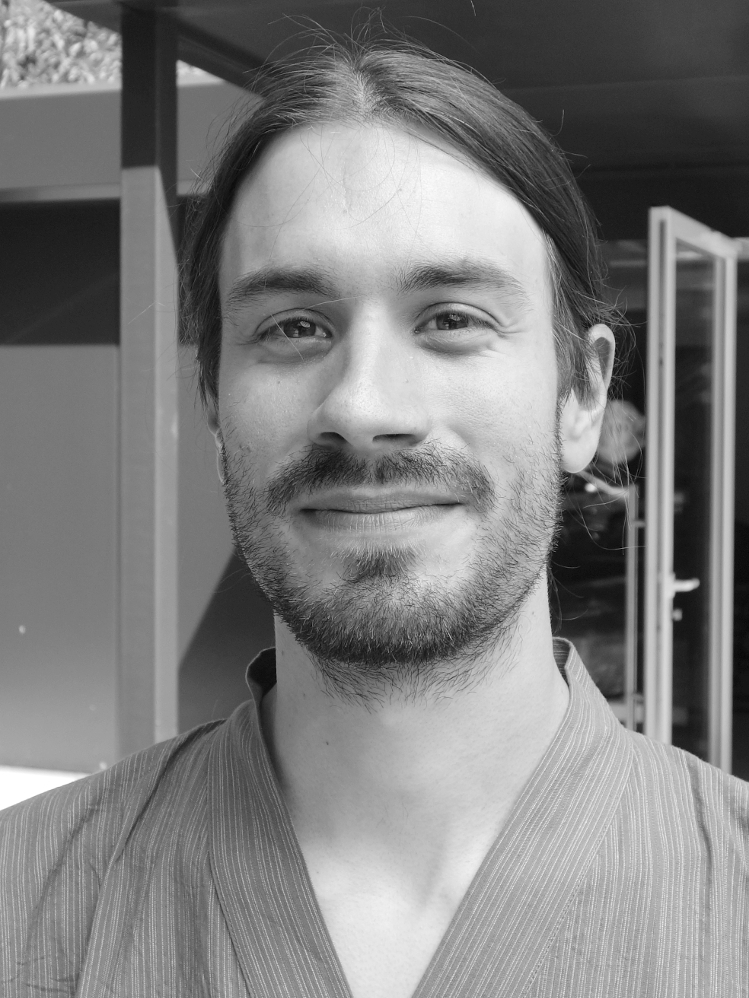}}]{Eric Müller }
	received the Ph.D.\ degree in physics from Heidelberg University, Germany, in 2014.
	Currently, he is a researcher in the Electronic Vision(s) group at Kirchhoff-Institute for Physics, Heidelberg University.
	His research interests are large-scale computing, information processing in closed-loop environments, and non-von-Neumann computing paradigms.
	He is the architect of the BrainScaleS operating system and leads BrainScaleS software development.
\end{IEEEbiography}

\begin{IEEEbiography}[{\includegraphics[width=1in,height=1.25in,clip,keepaspectratio]{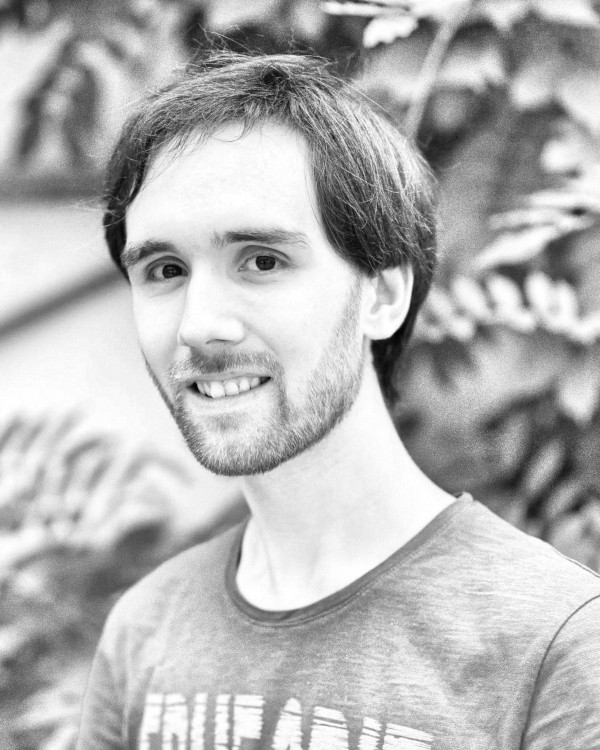}}]{Philipp Spilger }
	received the M.Sc.\ degree in physics from Heidelberg University, Germany, in 2021.
	Currently, he is a Ph.D.\ student in the Electronic Vision(s) group at Kirchhoff-Institute for Physics, Heidelberg University.
	His research interests are software abstraction for control and configuration of neuromorphic hardware based on optimization and compilation techniques.
\end{IEEEbiography}

\begin{IEEEbiography}[{\includegraphics[width=1in,height=1.25in,clip,keepaspectratio]{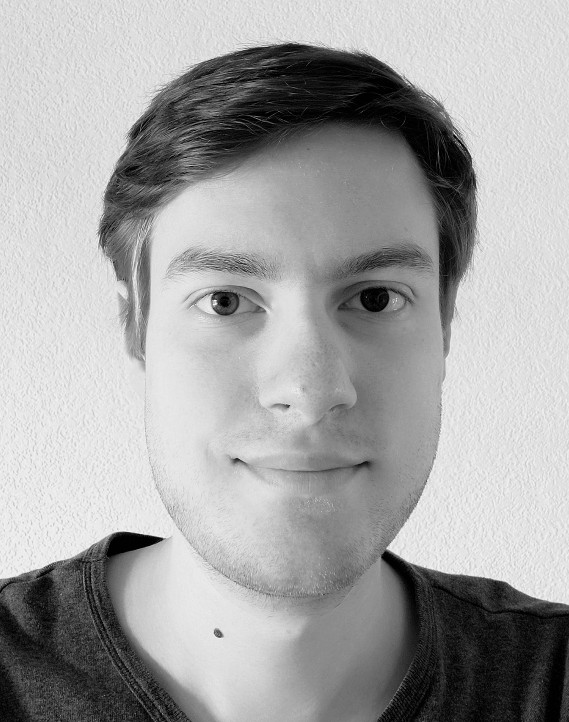}}]{Johannes Weis }
	received the M.Sc.\ degree in physics from Heidelberg University, Germany, in 2020.
    Currently, he is a researcher in the Electronic Vision(s) group at Kirchhoff-Institute for Physics, Heidelberg University.
    His research interests are characterization and calibration of analog VSLI systems for emulation of biologically inspred neural networks, including hardware-specific model development and optimization.
\end{IEEEbiography}

\begin{IEEEbiography}[{\includegraphics[width=1in,height=1.25in,clip,keepaspectratio]{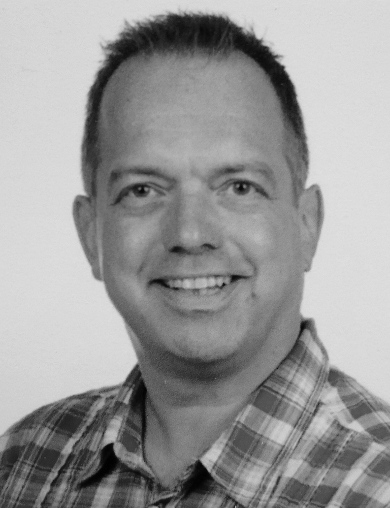}}]{Johannes Schemmel }
(M’08)
	received the Ph.D.\ degree in physics from Heidelberg University, Germany, in 1999.
	Currently, he is `Akademischer Oberrat' in the Kirchhoff Institute of Physics, Heidelberg, where he is head of the ASIC lab and the Electronic Vision(s) group.
	His research interests are mixed-mode VLSI systems for information processing, especially the analog implementation of biologically realistic neural network models.
	He is the architect of the Spikey and BrainScaleS accelerated Neuromorphic hardware systems.
\end{IEEEbiography}

\end{document}